\documentclass{article}

\usepackage{arxiv}

\usepackage[utf8]{inputenc} 
\usepackage[T1]{fontenc}    
\usepackage{hyperref}       
\usepackage{url}            
\usepackage{booktabs}       
\usepackage{amsfonts}       
\usepackage{nicefrac}       
\usepackage{microtype}      
\usepackage{lipsum}
\usepackage{graphicx}
\usepackage{epsfig}
\usepackage{subcaption}
\usepackage{amsmath,amssymb} 

\title{Real Image Super Resolution via Heterogeneous Model Ensemble using GP-NAS}

\author{
  Zhihong Pan, Baopu Li \\
  Baidu Research (USA)\\
   \\
   \And
 Teng Xi, Yanwen Fan, Gang Zhang, Jingtuo Liu, Junyu Han, Errui Ding \\
  Department of Computer Vision Technology (VIS), Baidu Inc.\\
  \\
}

\begin{document}
\newcommand{\etal}{\textit{et al}.}

\maketitle

\begin{abstract}
With advancement in deep neural network (DNN), 
recent state-of-the-art (SOTA) image super-resolution (SR) methods have
achieved impressive performance using deep residual network with dense skip connections.
While these models perform well on benchmark dataset where
low-resolution (LR) images are constructed from high-resolution (HR) references with
known blur kernel, real image SR is more challenging when both images in the LR-HR pair
are collected from real cameras.
Based on existing dense residual networks, a Gaussian process based neural architecture search
(GP-NAS) scheme is utilized to find candidate network architectures
using a large search space by varying the number of dense residual blocks,
the block size and the number of features.  A suite of heterogeneous
models with diverse network structure and hyperparameter
are selected for model-ensemble to achieve outstanding performance in real image SR.
The proposed method won the first place in all three tracks of the
AIM 2020 Real Image Super-Resolution Challenge.
\end{abstract}


\section{Introduction}
\label{sec:intro}

Image super-resolution (SR) refers the process to recover high-resolution (HR) 
images from low-resolution (LR) inputs. It is an important image
processing technique to enhance image quality which
subsequently helps to improve higher-level computer vision
tasks~\cite{dai_wacv_2016} \cite{haris_arxiv_2018}.
Over the years, many classical SR methods have been
proposed to successfully use various levels
of features like statistics~\cite{pickup_cj_2009},
edges~\cite{sun_cvpr_2008} \cite{tai_cvpr_2010}
and patches~\cite{wang_iccv_2005} \cite{yang_cvpr_2008} to restore HR images from LR inputs.
While there are also methods developed for SR
using multiple frames~\cite{khattab_iccic_2018},
the scope of introduction here is limited to
single image super-resolution (SISR).

More recently, the powerful deep learning
techniques have led to developments of 
many deep learning based SR models
\cite{dong_eccv_2014} \cite{kim_cvpr_2016_2} \cite{ledig_cvpr_2017} \cite{lim_cvprw_2017} \cite{zhang_cvpr_2018} \cite{zhang_eccv_2018}.
These deep learning models commonly rely on a large set of synthetic
training image pairs, where the LR input is downsampled from
the HR reference image using bicubic 
interpolation with antialiasing filters.  Common image quality metrics
used to assess performance of these SR models include peak 
signal-to-noise ratio (PSNR) and the structural 
similarity index (SSIM) \cite{wang_tip_2004},
both emphasizing image restoration fidelity by comparing to the HR reference. 
This may lead to SR results of high PSNR values but lack
of HR details perceptually.
Lately, a new metric LPIPS \cite{zhang_r_cvpr_2018}
is proposed to apply image features extracted from pretrained
Alexnet \cite{krizhevsky_nips_2012} to compare two images.
The smaller LPIPS is, the closer the generated SR image is to
the HR reference perceptually. 
With advancements in Generative Adversarial Nets (GAN)
\cite{goodfellow_nips_2014}, SR models trained using GAN
\cite{wang_eccv_2018} \cite{ma_cvpr_2020} \cite{ji_cvpr_2020}
have achieved the best performance of image perceptual
quality as compared to LPIPS.

In the past few years, neural architecture search (NAS) that aims to find the optimal network 
structure has received a lot of attention\cite{liu_arxiv_2017} 
\cite{zoph_arxiv_2016} \cite{zoph_cvpr_2018} \cite{liu_eccv_2018}. It effectively boosts the 
SOTA in many typical computer vision problems such as image classification\cite{liu_arxiv_2018}, 
object detection\cite{liang_iclr_2020}, segmentation\cite{chen_pami_2017} and so on. 
Most recently, some researches also begin to apply NAS for image SR 
problems \cite{chu_xx_arxiv_2019} \cite{song_dh_aaai_2020} \cite{guo_y_arxiv_2020} with impressive results using
efficient SR models.

In general, the SR problem is ill-posed as there are
multiple HR images corresponding to a single LR image even
when the LR image is constructed from the HR reference using bicubic interpolation
without added noise.
This ambiguity increases when the blur kernel and noise statistics of the LR
are not known, and is even more prominent
in real image SR problems where the LR image is not constructed
from the HR reference.  With the increased uncertainty,
it is common to see different deep learning SR models lead
to different versions of the
restored HR images for a single LR image, especially
when the network architectures are quite different.
To achieve the best performance of the real image SR problem
set forth by the AIM 2020 challenge,
a new fusion scheme is proposed in this study to
generate the final SR output using multi-level ensemble from a suite of
heterogeneous deep learning models that are obtained by applying NAS approach. 
The main contributions of the proposed method include:

\begin{itemize}

\item[$\bullet$] A Gaussian Process based NAS (GP-NAS) is first utilized for super-resolution 
  with specially designed search space, which can efficiently search and obtain the key 
  architecture related parameters and can yield multiple candidate models.

\item[$\bullet$] A multi-level ensemble scheme is proposed in testing, including self-ensemble 
  for patches, as well as patch-ensemble and model-ensemble for full-size images.

\item[$\bullet$] The proposed method was applied for the AIM 2020 Real Image Super-Resolution Challenge
  and won the first place in all three tracks (upscaling factors of $\times$2, $\times$3
  and $\times$4) with a comfortable margin in both PSNR and SSIM.

\end{itemize}

\section{Related Works}
\label{sec:rwork}

\noindent\textbf{Deep Learning for Single Image Super-Resolution.} As 
the first successful application, Dong \etal~\cite{dong_eccv_2014}
proposed a deep CNN model for end-to-end LR to HR mapping and showed
that the training of the neural network is equivalent to global
optimization of traditional sparse-coding-based SR methods.
Kim \etal \cite{kim_cvpr_2016_2} designed a deeply recursive
neural network to raise SR performance without increasing
parameters for additional convolutions.  Ledig \etal \cite{ledig_cvpr_2017}
were the first to use GAN for SR, introducing a perceptual
loss function to generate photo-realistic SR images from LR inputs.
Inspired by other SR models using deep residual networks
\cite{kim_cvpr_2016_1} \cite{kim_cvpr_2016_2},
Lim \etal \cite{lim_cvprw_2017} simplified the network structure
by removing BN layers and optimized the training process to 
achieve the best restoration fidelity at that time.
Zhang \etal first applied dense skip-connections \cite{zhang_cvpr_2018}
and later channel attention module \cite{zhang_eccv_2018}
in deep residual network for further advancing of SOTA.
Most recently, Guo \etal \cite{guo_cvpr_2020} proposed
a dual-regression method by adding a second downsampling
model and corresponding loss to make sure the restored SR
image can best match the LR input after downsampled
by the co-trained secondary model.

\noindent\textbf{NAS for Single Image Super-Resolution.} As the first attempt to apply NAS for SR, 
Chu \etal \cite{chu_xx_arxiv_2019} made use of an elastic search method on both micro and macro 
level with a hybrid controller that profits from evolutionary computation and reinforcement 
learning (RL), achieving comparable performance of PSNR with light model.  Based on different types 
of residual blocks and evolutionary algorithm, Song \etal \cite{song_dh_aaai_2020} proposed a search 
method for better and more efficient network for image SR. Guo \etal \cite{guo_y_arxiv_2020} put 
forward a novel hierarchical NAS approach that considers both the cell-level and network-level 
design based on a RL controller.  While all the three works are promising at searching for 
efficient SR models where resources like model size or FLOPS are limited,
they are not able to achieve the high PSNR or SSIM values comparing to
other SOTA methods using manually designed residual networks with
dense skip connections.
Aiming at AIM 2020 challenge that does not take model efficiency in consideration, we 
mainly concentrate on the macro level structure design for a better network structure 
that can achieve the best SR performance in terms of PSNR and SSIM. Moreover, instead of 
using RL or evolutionary based search method that tend to be very time consuming, we apply 
GP-NAS approach to search the key network structure parameters such as the number of dense 
residual block, the block size and the number of features.      

\section{Problem Formulation}
\label{sec:pf}

Learning based image SR methods often rely on a large number of image
pairs, including low-res image $\mathbf{I}_{LR}$ and reference high-res
$\mathbf{I}_{HR}$.  For real image SR, as shown in Eq.~\ref{eq:ds},
$\mathbf{I}_{LR}$ could be modeled from $\mathbf{I}_{HR}$ using three steps:
convolution with a kernel $\mathbf{k}$, downsampling $D_s(\cdot)$ 
and addition of noise $\mathbf{n}$.

\begin{equation}
     \mathbf{I}_{LR}=D_s(\mathbf{I}_{HR}*\mathbf{k})+\mathbf{n}
\label{eq:ds}
\end{equation}

The goal of image SR is to reverse this process, finding the matching $\mathbf{I}_{HR}$ from a known $\mathbf{I}_{LR}$.
This problem is challenging as there are many versions of
$\mathbf{I}_{HR}$ that could generate the same $\mathbf{I}_{LR}$
following the process in Eq.~\ref{eq:ds}, even when the 
kernel $\mathbf{k}$ is known and there is no noise
$\mathbf{n}$.  Learning based SR model $f(\cdot)$ use a total of $n$
image pairs to minimize the average error as in Eq.~\ref{eq:sr}.
It is common that only $\mathbf{I}^i_{HR}$ is a real image
and the corresponding $\mathbf{I}^i_{LR}$ is constructed from
$\mathbf{I}^i_{HR}$ following the process described in Eq.~\ref{eq:ds}.

\begin{equation}
     \arg\min_{f} \sum \|f(\mathbf{I}^i_{LR})-\mathbf{I}^i_{HR} \|, 
     i \in \{1, 2, \cdots, n\}
\label{eq:sr}
\end{equation}

In the case that both $\mathbf{I}_{HR}$ and $\mathbf{I}_{LR}$
are real images collected separately, the relationship between the image pair
is much more complicated.  To get a digital image $\mathbf{I}$ from
an object $\mathbf{O}$, there are three transformations included in general
as shown in Eq.~\ref{eq:imaging}.  The optical transformation
$\mathbb{O}(\cdot)$ refers to the process of photons reflected
from the object passing through the lens of the camera.  Optical
characteristics of the lens like modulation transfer function (MTF)
and lens settings like aperture are key variables here.  The second
transformation $\mathbb{D}(\cdot)$ refers to the analog to digital
converter (ADC) that turns photons to digital numbers, where
the noise is introduced.  The last one
$\mathbb{I}(\cdot)$ refers to the image signal processor (ISP)
which transforms noisy raw images to end result of sRGB images.
This step is the most complicated of three,
including multiple processes like denoising and color balancing
at both global and local levels.

\begin{equation}
     \mathbf{I} = \mathbb{I}(\mathbb{D}(\mathbb{O}(\mathbf{O})))
\label{eq:imaging}
\end{equation}

As each of the three transformations could be different for HR
and LR images, the relationship between $\mathbf{I}_{LR}$
and $\mathbf{I}_{HR}$ can be illustrated as in
Equations~\ref{eq:lr}-\ref{eq:lrhr}
where they are linked indirectly by the downsampling
of $\mathbf{O}_{HR}$ to $\mathbf{O}_{LR}$ using $D_s(\cdot)$.
With all these added variations, the real image SR becomes
more challenging.  For example, $\mathbf{I}_{HR}$ is
more clear in general compared to $\mathbf{I}_{LR}$.
But for background objects at a further distance,
they could be more blurry in $\mathbf{I}_{HR}$ if its lens
has a smaller f-number which lead to smaller depth-of-field.
The motivation to use heterogeneous model ensemble is based on
the observation that different model could lead to optimization results
biased towards different variation factors
even when using the same set of training image pairs.

\begin{align}
    & \mathbf{I}_{LR} = \mathbb{I}_{LR}(\mathbb{D}_{LR}(\mathbb{O}_{LR}(\mathbf{O}_{LR}))) \label{eq:lr} \\
    & \mathbf{I}_{HR} = \mathbb{I}_{HR}(\mathbb{D}_{HR}(\mathbb{O}_{HR}(\mathbf{O}_{HR}))) \label{eq:hr} \\
    & \mathbf{O}_{LR} = D_s(\mathbf{O}_{HR}) \label{eq:lrhr}
\end{align}

\section{Proposed Real Image SR Method}
\label{sec:method}

The proposed real image SR method using dense residual network, GP-NAS and heterogeneous model ensemble
is explained in this section.  First, the primary dense residual network
and the search space of different hyperparameters are introduced.
Then, method to find heterogeneous models using GP-NAS is explained, followed by 
the multi-level ensemble that is used to generate full-size SR images for the AIM 2020 challenge.

\subsection{Dense Residual Network (DRN)}
\label{sec:drn}

 \begin{figure}
 \begin{center}
     \includegraphics[width=\linewidth]{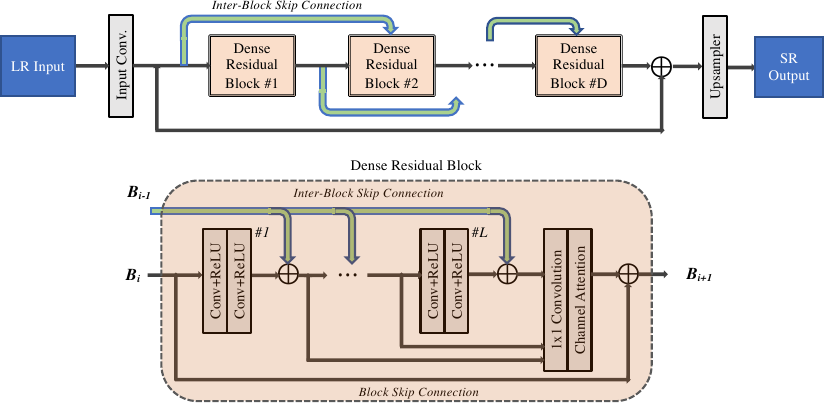}
 \end{center}
 \caption{The deep dense residual network architecture for image super resolution.}
 \vspace{-10pt}
 \label{fig:drn}
 \end{figure}

The backbone model of the proposed method is 
a deep dense residual network originally developed for raw image demosaicking and denoising.
As depicted in Fig.~\ref{fig:drn},
in addition to the shallow feature convolution at the front and the upsampler at the end, the proposed 
network consists of a total depth of $D$ dense residual blocks (DRB).  The input convolution layer 
converts the $3$-channel LR input to a total of $F$-channel shallow features.
For the middle DRB blocks, each one includes $L$ stages
of double layers of convolution and the outputs of all $L$ stages are concatenated together
before convoluted from $F\times L$ to $F$ channels.  An additional channel-attention layeris 
included at the end of each block, similar to RCAN~\cite{zhang_eccv_2018}.  There are two types of skip connections
included in each block, the block skip connection (BSC) and inter-block skip connection (IBSC).
The BSC is the shortcut between input and output of block $B_i$, while the IBSC includes
two shortcuts from the input of block $B_{i-1}$ to the two stages inside block $B_i$ respectively.  The 
various skip connections, especially IBSC, are included to combine features with a large range of receptive fields.
The last block is an enhanced upsampler that transforms all $F$-channel LR features
to the estimated $3$-channel SR image.
This dense residual network has three main hyperparameters:
$F$ is the number of feature channels, $D$ is the number of DRB layers
and $L$ is the number of stages for each DRB. All these three hyperparameters will greatly affect 
the performance of SR.  Previous efforts mainly use professional expertise or experience to choose 
them based on, which is laborious. To overcome this issue, we apply NAS to search for 
the optimal network structure, which will be elaborated in the subsequent subsection.

\subsection{Gaussian Process based Neural Architecture Search}
\label{sec:nas}

\begin{figure}[ht!]
 \begin{center}
     \includegraphics[width=\linewidth]{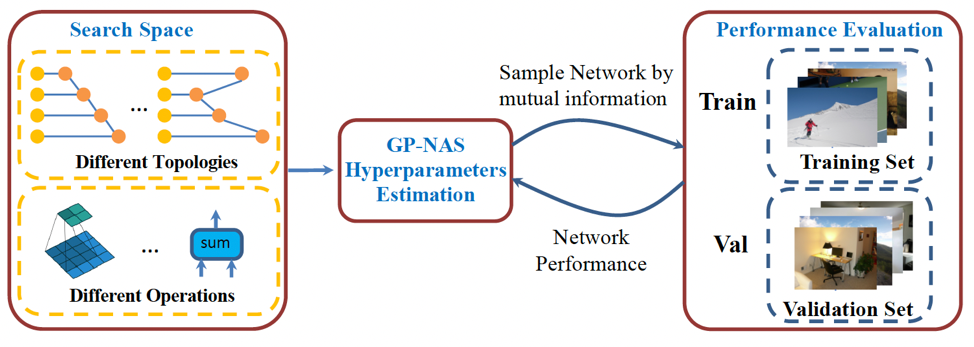}
 \end{center}
 \caption{The framework of the GP-NAS.}
 \vspace{-10pt}
 \label{fig:gp-nas}
 \end{figure}

Since most NAS methods are still time consuming, we had proposed Gaussian Process based Neural 
Architecture Search (GP-NAS) \cite{li_cvpr_2020} to accelerate the searching process.
Figure \ref{fig:gp-nas} illustrates the framework of the GP-NAS. The GP-NAS formulates NAS 
from a Bayesian perspective. Specifically, given the hyper-parameters of GP-NAS, we are capable of 
predicting the performance of any architectures in the search space effectively. Then, the NAS 
process is converted to hyperparameters estimation. By mutual information maximization, we can 
efficiently sample networks. Accordingly, based on the performances of sampled networks,
the posterior distribution of hyperparameters can be gradually and efficiently updated. 
Based on the estimated hyperparameters, the architecture with best performance can be obtained. 
More details about our proposed GP-NAS can be found in \cite{li_cvpr_2020}. 

\subsection{Multi-Level Ensemble}
\label{sec:mle}

Targeting for the AIM 2020 challenge, where the test images are much larger than the training patches, a 
multi-levelensemble scheme is designed to achieve optimal image restoration quality. First, for model-ensemble, 
the input LR image is processed by a suite of heterogeneous models separately
and the output HR images are averaged to get the final output.  Additionally, each full size LR input is 
cropped to patches, with each has an overlapping buffer with neighbouring patches.  A patch-ensemble method 
is then used to blend all restored HR patches together, using different weights for each pixel which are 
correlated to the distance between the patch center and corresponding pixels.  The most commonly used 
self-ensemble is also applied by flipping and/or transposing the input patch before restoration.

\begin{table}[t!]
 \small
 \begin{center}
  \begin{tabular}{rcccccc}
  \toprule
      {} & \multicolumn{2}{c}{Upscaling $\times 2$} & \multicolumn{2}{c}{Upscaling $\times 3$} & \multicolumn{2}{c}{Upscaling $\times 4$} \\
      {} & {PSNR} & {SSIM} & {PSNR} & {SSIM} & {PSNR} & {SSIM} \\
 \midrule
 DRN\textsuperscript{*} & 32.51\textsuperscript{1} & 0.9209\textsuperscript{1} & 31.07\textsuperscript{1} & 0.8796\textsuperscript{1} & 30.26\textsuperscript{3} & 0.                     8401\textsuperscript{3} \\
 RCAN\textsuperscript{*} & 32.31\textsuperscript{3} & 0.9188\textsuperscript{3} & 30.97\textsuperscript{3} & 0.8787\textsuperscript{3} & 30.31\textsuperscript{2} & 0.                    8403\textsuperscript{1} \\
 RCAN & 32.40\textsuperscript{2} & 0.9200\textsuperscript{2} & 31.01\textsuperscript{2} & 0.8792\textsuperscript{2} & 30.32\textsuperscript{1} & 0.8402\textsuperscript{1} \\
 \midrule
 DRN\textsuperscript{*}+RCAN\textsuperscript{*} & 32.56\textsuperscript{2} & 0.9210\textsuperscript{2} & 31.21\textsuperscript{2} & 0.8811\textsuperscript{3} & 30.47\textsuperscript{3}  & 0.8423\textsuperscript{1} \\
 RCAN\textsuperscript{*}+RCAN & 32.49\textsuperscript{3} & 0.9204\textsuperscript{3} & 31.17\textsuperscript{3} & 0.8807\textsuperscript{3} & 30.48\textsuperscript{1} & 0.8421\textsuperscript{3} \\
 DRN\textsuperscript{*}+RCAN & 32.62\textsuperscript{1} & 0.9215\textsuperscript{1} & 31.24\textsuperscript{1} & 0.8814\textsuperscript{1} & 30.48\textsuperscript{1} & 0.8423\textsuperscript{1} \\
 \midrule
 3-Model Ensemble & \textbf{32.63} & \textbf{0.9218} & \textbf{31.28} & \textbf{0.8822} & \textbf{30.55} & \textbf{0.8435} \\
  \bottomrule
  \end{tabular}
 \end{center}
      \caption{Quantitative results of single-model and model-ensemble methods. Best results are in bold and the ranks in each category are superscripted.}
      \vspace{-10pt}
  \label{tab:as}
 \end{table}

\section{Experimental Results}

The AIM 2020 challenge aims to find a generic model to super-resolve LR images
captured in practical scenarios.  To achieve this goal, paired LR and HR images were 
taken by various DSLR cameras.  However, images used for training, validation and testing are 
captured in the same way with the same set of cameras, so the transformation processes described in
Equations~\ref{eq:lr}-\ref{eq:lrhr} are not changed among images
of the same upscaling factor, meaning the learned transformation from training data
is expected to achieve similar results on the validation and test images.
For this new real image SR dataset,
a total of $19,000$ LR-HR pairs are available for model training for each
of the $\times 2$, $\times 3$ and $\times 4$ upscaling factors.
The LR image resolutions are $380\times 380$ for $\times 3$, $272\times 272$ for
$\times 2$ and $194\times 194$ for $\times 4$ respectively.

For our experiments of each upscaling factor, $600$ of the $19,000$ pairs
are reserved for validation while the remaining ones are used for training.
Note that any LR-HR pairs that are not perfectly aligned,
those with normalized cross-correlation (NCC) less than 0.99,
were excluded from both training and validation.
For each epoch, a $120\times 120$ patch is randomly cropped and augmented
with flipping and transposing from each training image.  A mixed loss of $L1$
and multi-scale structural similarity (MS-SSIM) is taken for training.
For the experiment, the new model candidate search scheme using
GP-NAS was implemented in PaddlePaddle~\cite{ma_fdd_2019}
and the final-training of searched models were conducted using
PyTorch~\cite{paszke_nips_2019}

\begin{figure}[hb!]
\captionsetup[subfigure]{labelformat=empty}
\begin{center}
  \begin{subfigure}[b]{0.16\textwidth}
    \centering
      \includegraphics[width=\textwidth, interpolate=false, clip=true]{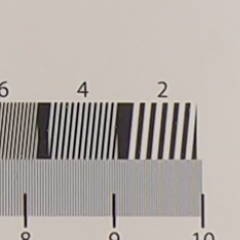}
  \end{subfigure}
  \begin{subfigure}[b]{0.16\textwidth}
    \centering
      \includegraphics[width=\textwidth, interpolate=false, clip=true]{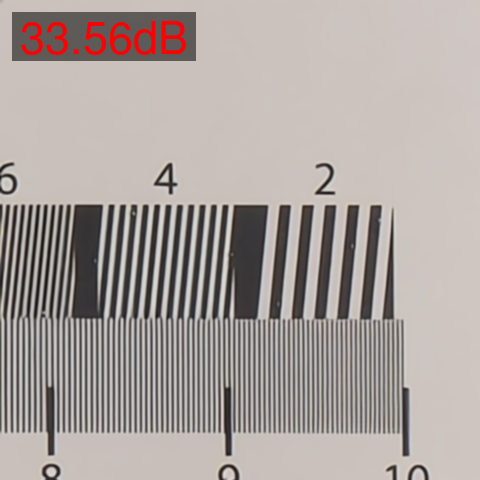}
  \end{subfigure}
  \begin{subfigure}[b]{0.16\textwidth}
    \centering
      \includegraphics[width=\textwidth, interpolate=false, clip=true]{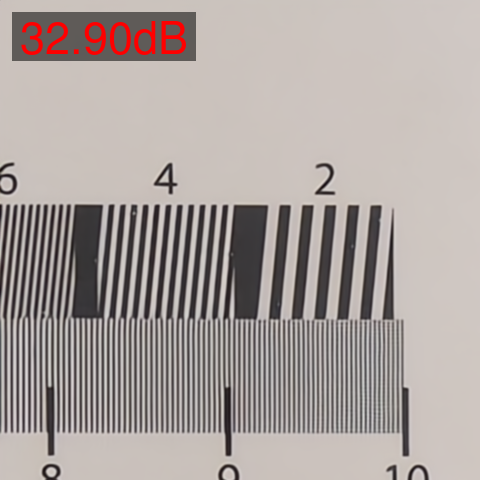}
  \end{subfigure}
  \begin{subfigure}[b]{0.16\textwidth}
    \centering
      \includegraphics[width=\textwidth, interpolate=false, clip=true]{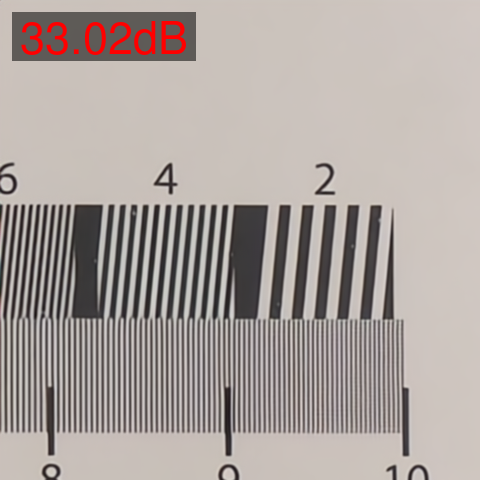}
  \end{subfigure}
  \begin{subfigure}[b]{0.16\textwidth}
    \centering
      \includegraphics[width=\textwidth, interpolate=false, clip=true]{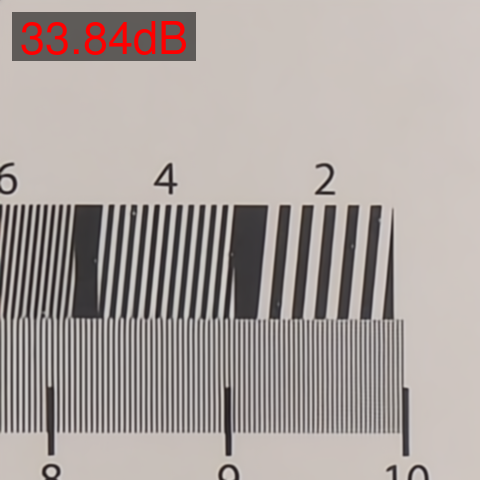}
  \end{subfigure}
  \begin{subfigure}[b]{0.16\textwidth}
    \centering
      \includegraphics[width=\textwidth, interpolate=false, clip=true]{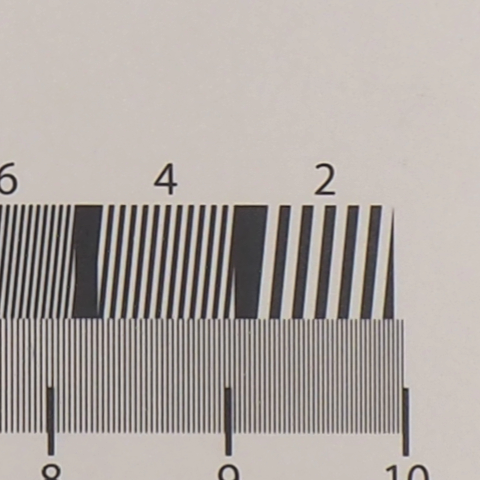}
  \end{subfigure}
  \begin{subfigure}[b]{0.16\textwidth}
    \centering
      \includegraphics[width=\textwidth, interpolate=false, clip=true]{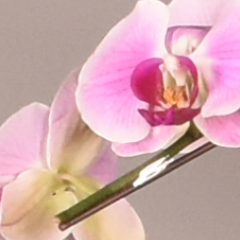}
  \end{subfigure}
  \begin{subfigure}[b]{0.16\textwidth}
    \centering
      \includegraphics[width=\textwidth, interpolate=false, clip=true]{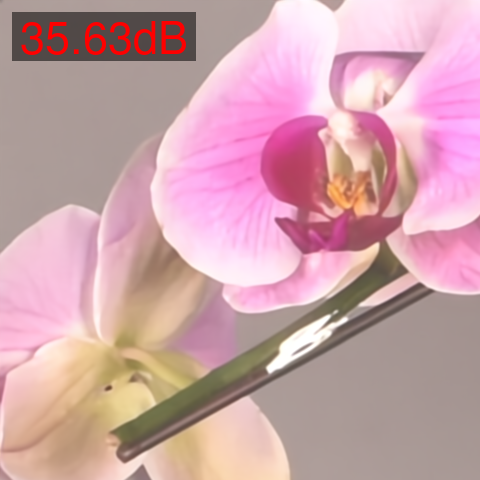}
  \end{subfigure}
  \begin{subfigure}[b]{0.16\textwidth}
    \centering
      \includegraphics[width=\textwidth, interpolate=false, clip=true]{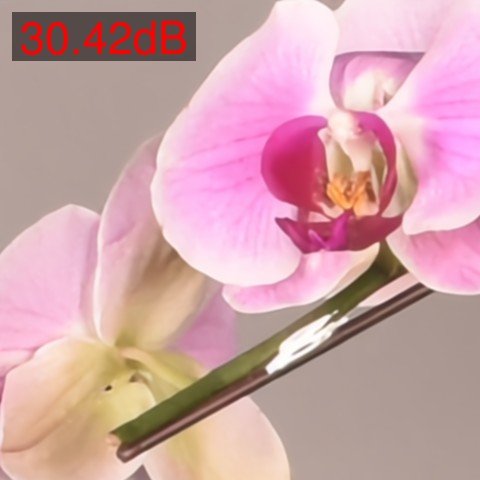}
  \end{subfigure}
  \begin{subfigure}[b]{0.16\textwidth}
    \centering
      \includegraphics[width=\textwidth, interpolate=false, clip=true]{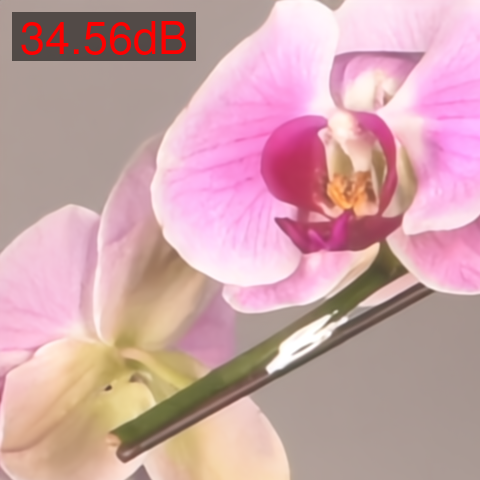}
  \end{subfigure}
  \begin{subfigure}[b]{0.16\textwidth}
    \centering
      \includegraphics[width=\textwidth, interpolate=false, clip=true]{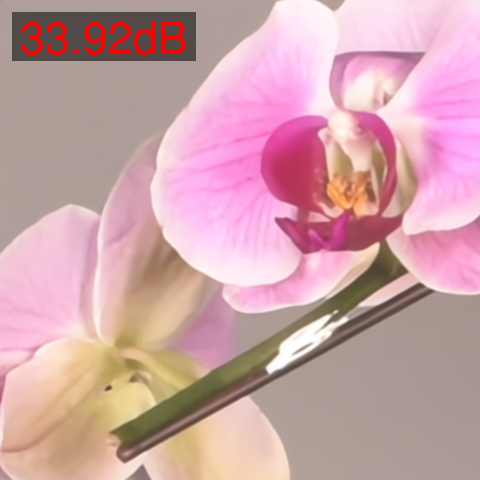}
  \end{subfigure}
  \begin{subfigure}[b]{0.16\textwidth}
    \centering
      \includegraphics[width=\textwidth, interpolate=false, clip=true]{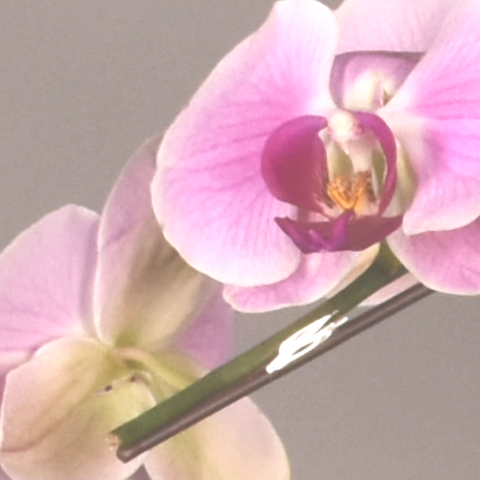}
  \end{subfigure}
  \begin{subfigure}[b]{0.16\textwidth}
    \centering
      \includegraphics[width=\textwidth, interpolate=false, clip=true]{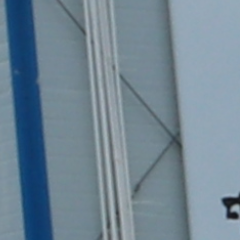}
      \caption{LR}
  \end{subfigure}
  \begin{subfigure}[b]{0.16\textwidth}
    \centering
      \includegraphics[width=\textwidth, interpolate=false, clip=true]{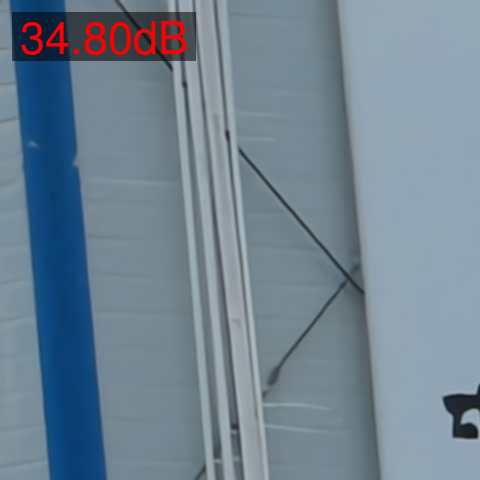}
      \caption{DRN\textsuperscript{*}}
  \end{subfigure}
  \begin{subfigure}[b]{0.16\textwidth}
    \centering
      \includegraphics[width=\textwidth, interpolate=false, clip=true]{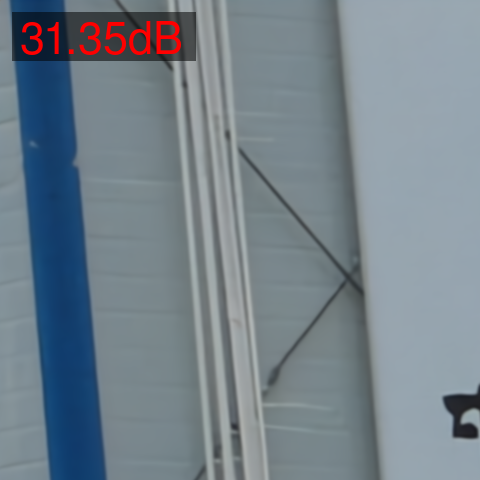}
      \caption{RCAN\textsuperscript{*}}
  \end{subfigure}
  \begin{subfigure}[b]{0.16\textwidth}
    \centering
      \includegraphics[width=\textwidth, interpolate=false, clip=true]{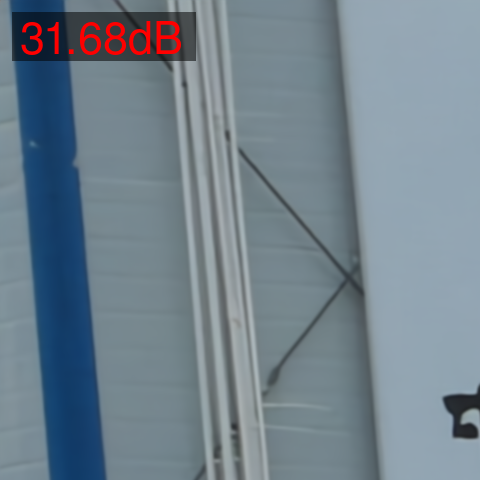}
      \caption{RCAN}
  \end{subfigure}
  \begin{subfigure}[b]{0.16\textwidth}
    \centering
      \includegraphics[width=\textwidth, interpolate=false, clip=true]{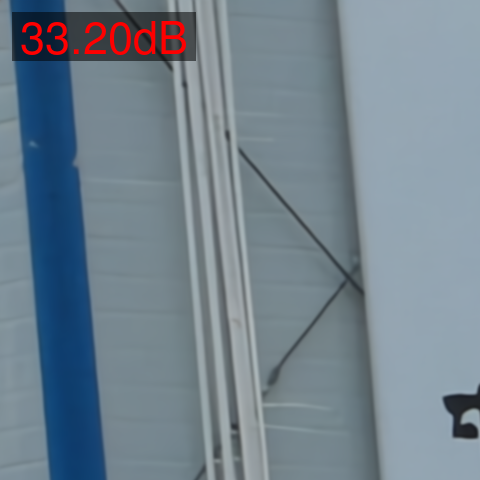}
      \caption{Ensemble}
  \end{subfigure}
  \begin{subfigure}[b]{0.16\textwidth}
    \centering
      \includegraphics[width=\textwidth, interpolate=false, clip=true]{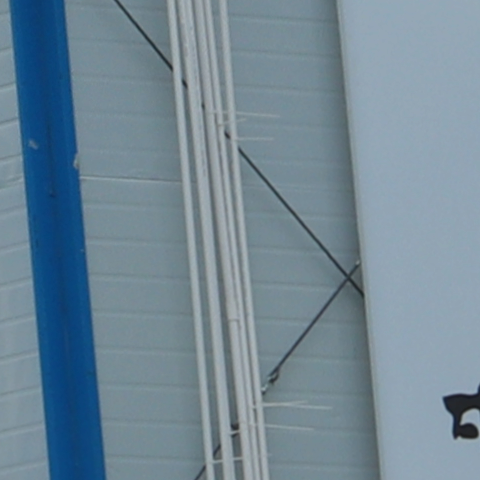}
      \caption{HR}
  \end{subfigure}
\end{center}
    \vspace{-10pt}
    \caption{Visual and quantitative comparison of $\times 2$ SR results.}
    \vspace{-10pt}
\label{fig:x2}
\end{figure}

\begin{figure}[ht!]
\captionsetup[subfigure]{labelformat=empty}
\begin{center}
  \begin{subfigure}[b]{0.16\textwidth}
    \centering
      \includegraphics[width=\textwidth, interpolate=false, clip=true]{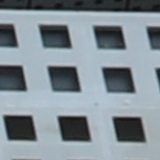}
  \end{subfigure}
  \begin{subfigure}[b]{0.16\textwidth}
    \centering
      \includegraphics[width=\textwidth, interpolate=false, clip=true]{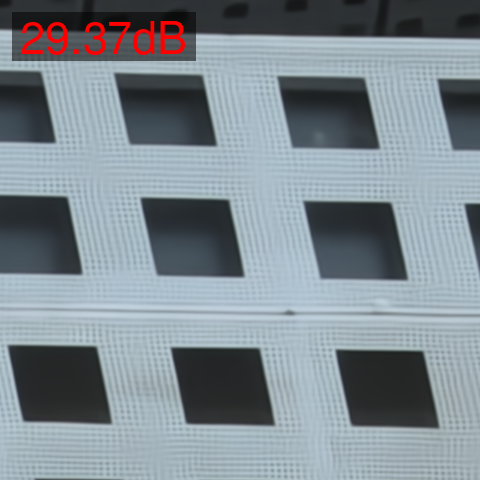}
  \end{subfigure}
  \begin{subfigure}[b]{0.16\textwidth}
    \centering
      \includegraphics[width=\textwidth, interpolate=false, clip=true]{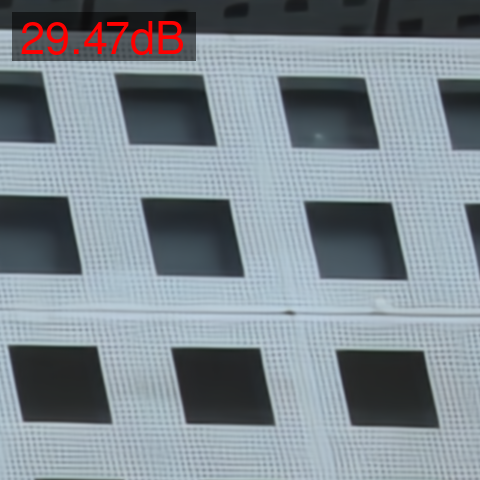}
  \end{subfigure}
  \begin{subfigure}[b]{0.16\textwidth}
    \centering
      \includegraphics[width=\textwidth, interpolate=false, clip=true]{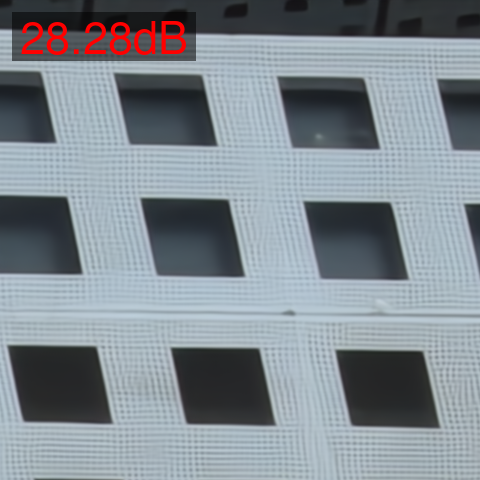}
  \end{subfigure}
  \begin{subfigure}[b]{0.16\textwidth}
    \centering
      \includegraphics[width=\textwidth, interpolate=false, clip=true]{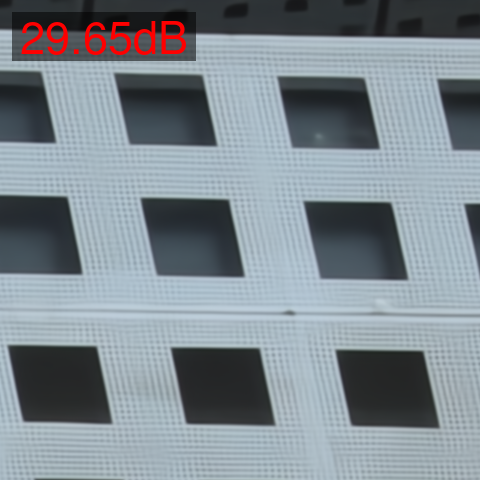}
  \end{subfigure}
  \begin{subfigure}[b]{0.16\textwidth}
    \centering
      \includegraphics[width=\textwidth, interpolate=false, clip=true]{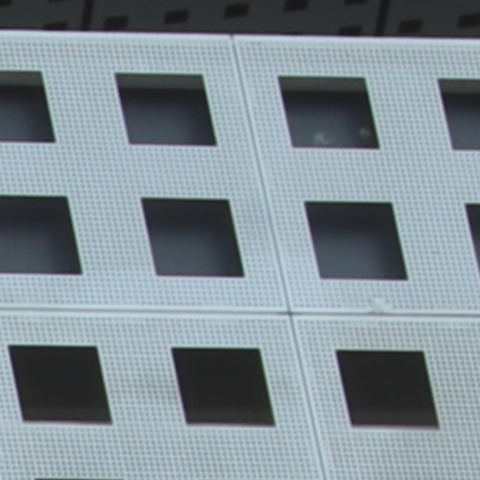}
  \end{subfigure}
  \begin{subfigure}[b]{0.16\textwidth}
    \centering
      \includegraphics[width=\textwidth, interpolate=false, clip=true]{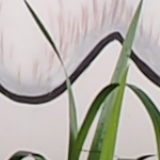}
  \end{subfigure}
  \begin{subfigure}[b]{0.16\textwidth}
    \centering
      \includegraphics[width=\textwidth, interpolate=false, clip=true]{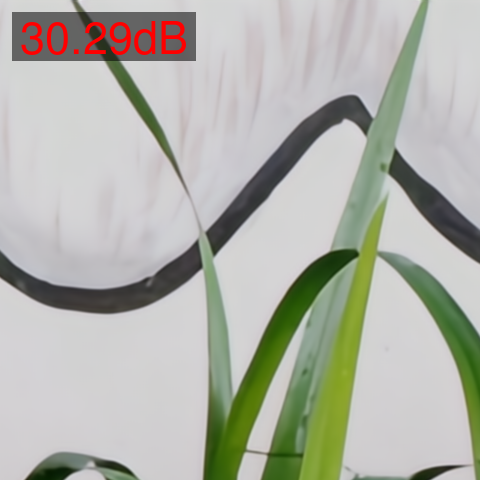}
  \end{subfigure}
  \begin{subfigure}[b]{0.16\textwidth}
    \centering
      \includegraphics[width=\textwidth, interpolate=false, clip=true]{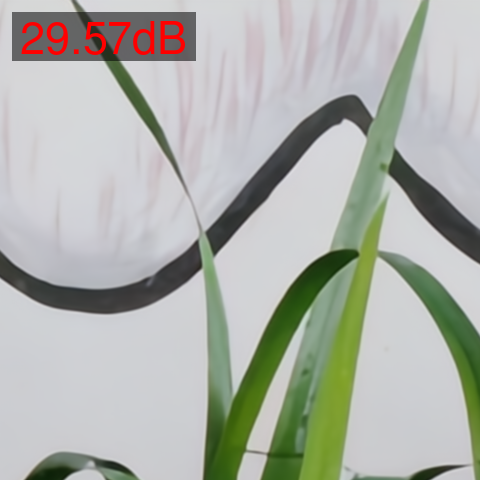}
  \end{subfigure}
  \begin{subfigure}[b]{0.16\textwidth}
    \centering
      \includegraphics[width=\textwidth, interpolate=false, clip=true]{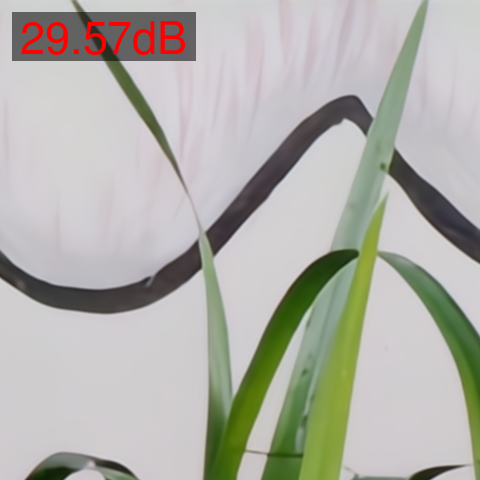}
  \end{subfigure}
  \begin{subfigure}[b]{0.16\textwidth}
    \centering
      \includegraphics[width=\textwidth, interpolate=false, clip=true]{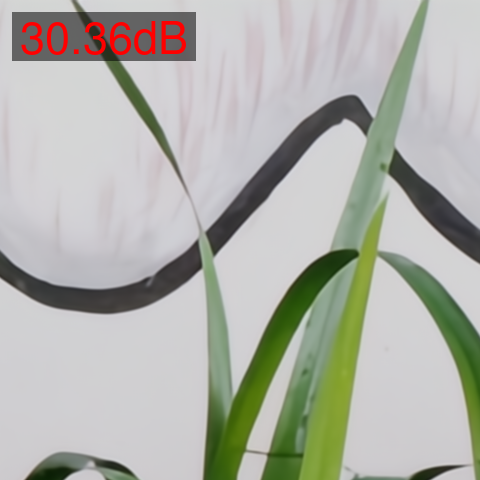}
  \end{subfigure}
  \begin{subfigure}[b]{0.16\textwidth}
    \centering
      \includegraphics[width=\textwidth, interpolate=false, clip=true]{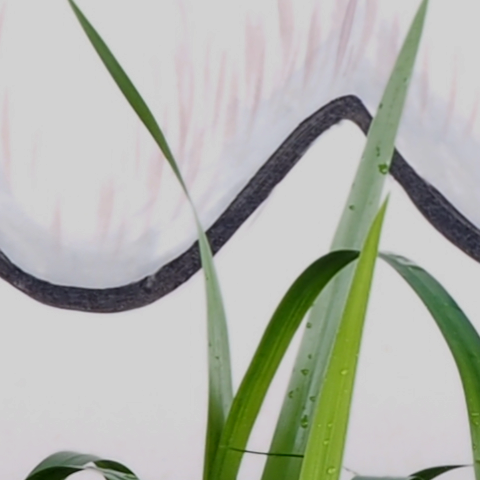}
  \end{subfigure}
  \begin{subfigure}[b]{0.16\textwidth}
    \centering
      \includegraphics[width=\textwidth, interpolate=false, clip=true]{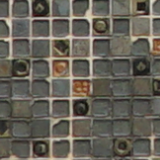}
      \caption{LR}
  \end{subfigure}
  \begin{subfigure}[b]{0.16\textwidth}
    \centering
      \includegraphics[width=\textwidth, interpolate=false, clip=true]{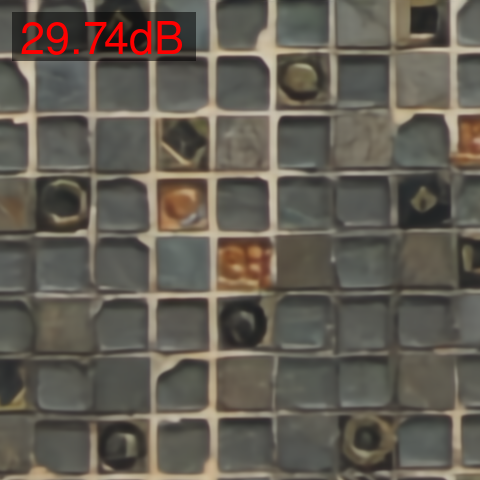}
      \caption{DRN\textsuperscript{*}}
  \end{subfigure}
  \begin{subfigure}[b]{0.16\textwidth}
    \centering
      \includegraphics[width=\textwidth, interpolate=false, clip=true]{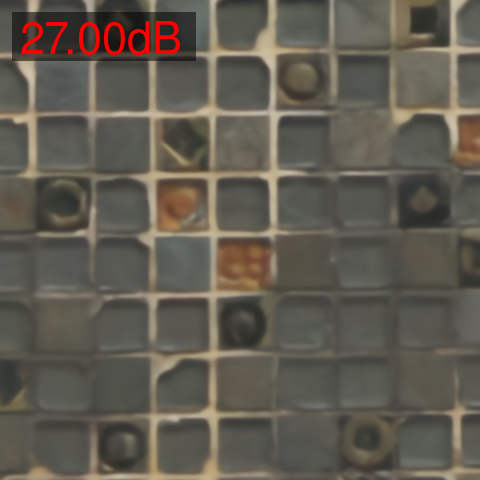}
      \caption{RCAN\textsuperscript{*}}
  \end{subfigure}
  \begin{subfigure}[b]{0.16\textwidth}
    \centering
      \includegraphics[width=\textwidth, interpolate=false, clip=true]{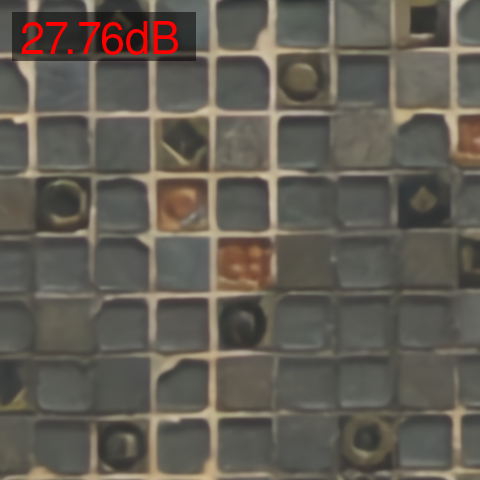}
      \caption{RCAN}
  \end{subfigure}
  \begin{subfigure}[b]{0.16\textwidth}
    \centering
      \includegraphics[width=\textwidth, interpolate=false, clip=true]{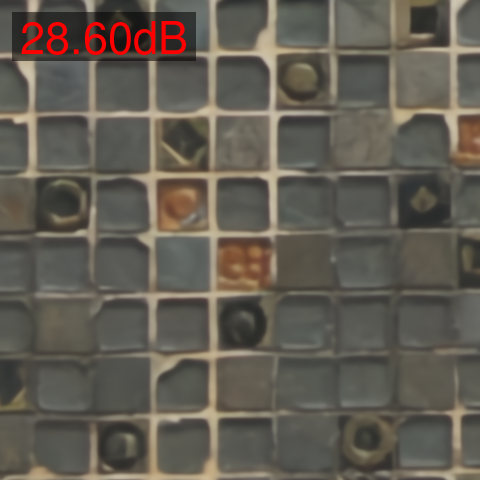}
      \caption{Ensemble}
  \end{subfigure}
  \begin{subfigure}[b]{0.16\textwidth}
    \centering
      \includegraphics[width=\textwidth, interpolate=false, clip=true]{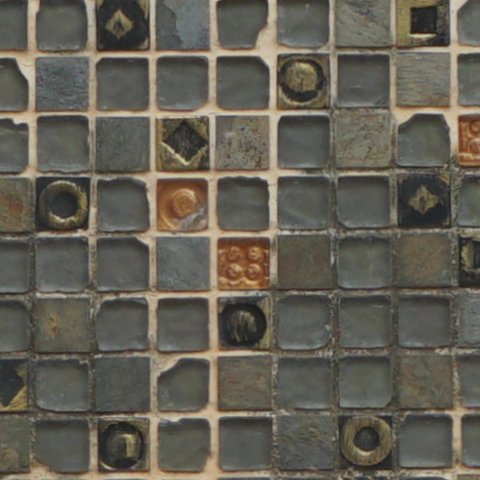}
      \caption{HR}
  \end{subfigure}
\end{center}
    \vspace{-10pt}
    \caption{Visual and quantitative  comparison of $\times 3$ SR results.}
    \vspace{-10pt}
\label{fig:x3}
\end{figure}

\begin{figure}[ht!]
\captionsetup[subfigure]{labelformat=empty}
\begin{center}
  \begin{subfigure}[b]{0.16\textwidth}
    \centering
      \includegraphics[width=\textwidth, interpolate=false, clip=true]{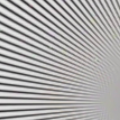}
  \end{subfigure}
  \begin{subfigure}[b]{0.16\textwidth}
    \centering
      \includegraphics[width=\textwidth, interpolate=false, clip=true]{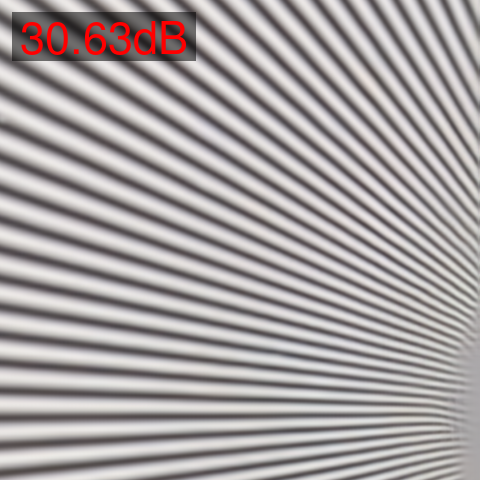}
  \end{subfigure}
  \begin{subfigure}[b]{0.16\textwidth}
    \centering
      \includegraphics[width=\textwidth, interpolate=false, clip=true]{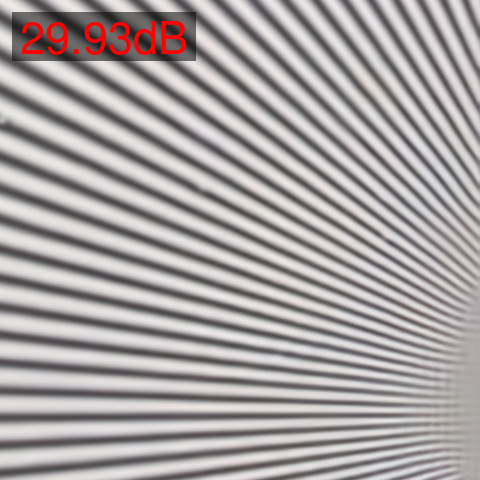}
  \end{subfigure}
  \begin{subfigure}[b]{0.16\textwidth}
    \centering
      \includegraphics[width=\textwidth, interpolate=false, clip=true]{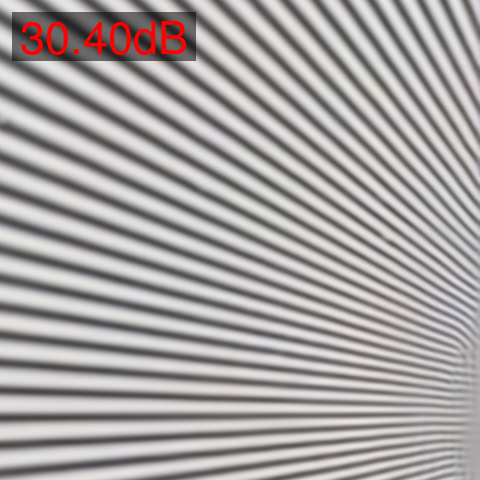}
  \end{subfigure}
  \begin{subfigure}[b]{0.16\textwidth}
    \centering
      \includegraphics[width=\textwidth, interpolate=false, clip=true]{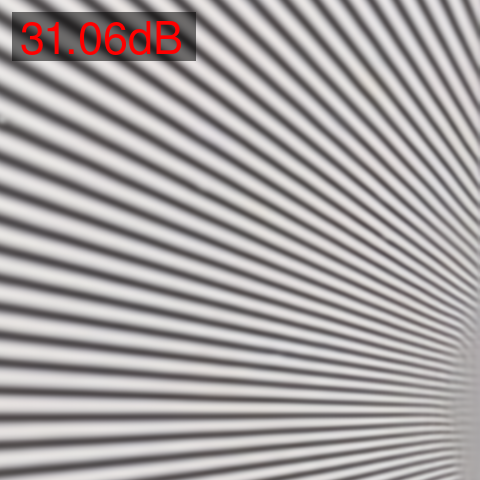}
  \end{subfigure}
  \begin{subfigure}[b]{0.16\textwidth}
    \centering
      \includegraphics[width=\textwidth, interpolate=false, clip=true]{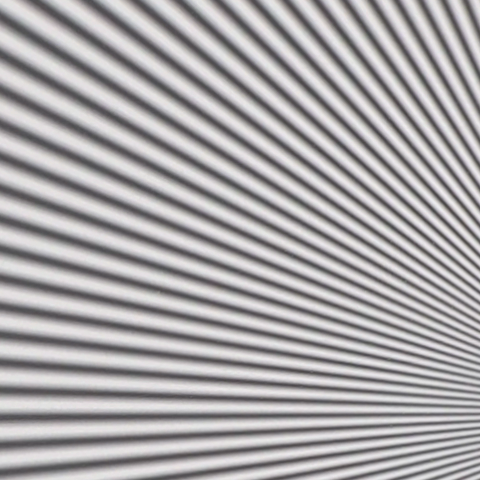}
  \end{subfigure}
  \begin{subfigure}[b]{0.16\textwidth}
    \centering
      \includegraphics[width=\textwidth, interpolate=false, clip=true]{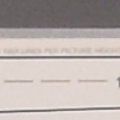}
  \end{subfigure}
  \begin{subfigure}[b]{0.16\textwidth}
    \centering
      \includegraphics[width=\textwidth, interpolate=false, clip=true]{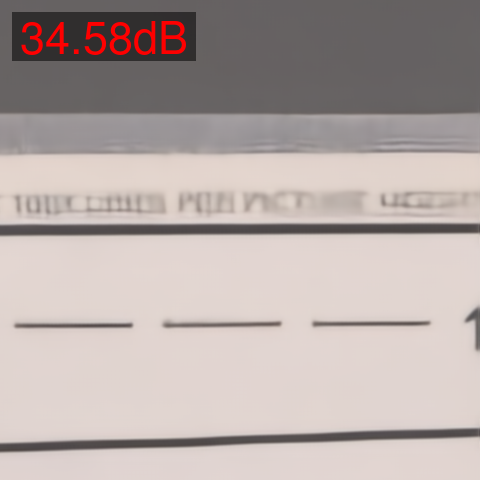}
  \end{subfigure}
  \begin{subfigure}[b]{0.16\textwidth}
    \centering
      \includegraphics[width=\textwidth, interpolate=false, clip=true]{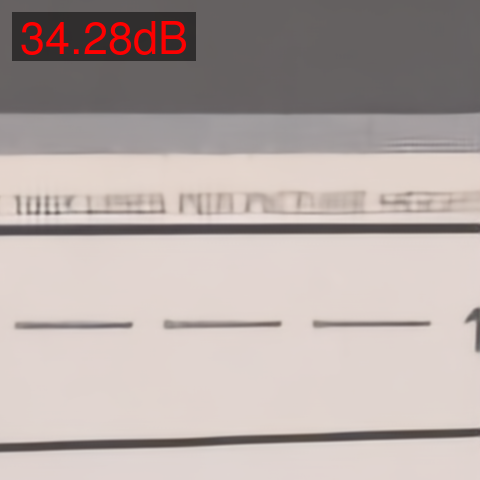}
  \end{subfigure}
  \begin{subfigure}[b]{0.16\textwidth}
    \centering
      \includegraphics[width=\textwidth, interpolate=false, clip=true]{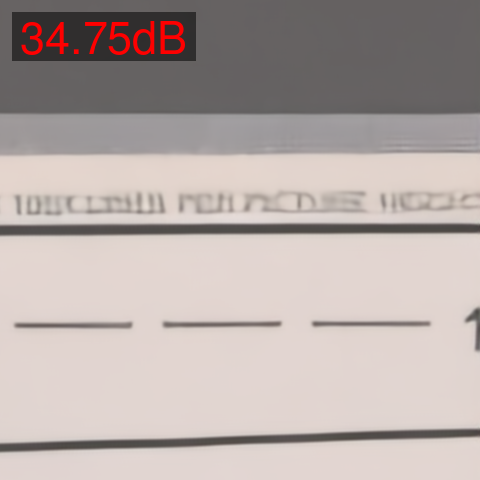}
  \end{subfigure}
  \begin{subfigure}[b]{0.16\textwidth}
    \centering
      \includegraphics[width=\textwidth, interpolate=false, clip=true]{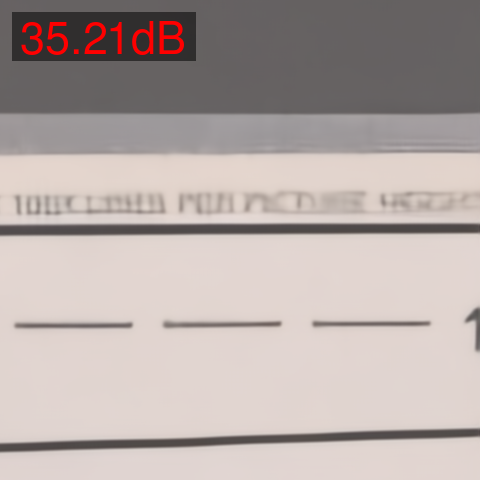}
  \end{subfigure}
  \begin{subfigure}[b]{0.16\textwidth}
    \centering
      \includegraphics[width=\textwidth, interpolate=false, clip=true]{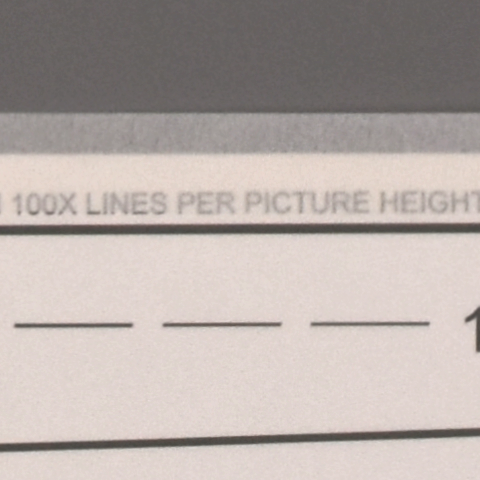}
  \end{subfigure}
  \begin{subfigure}[b]{0.16\textwidth}
    \centering
      \includegraphics[width=\textwidth, interpolate=false, clip=true]{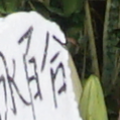}
      \caption{LR}
  \end{subfigure}
  \begin{subfigure}[b]{0.16\textwidth}
    \centering
      \includegraphics[width=\textwidth, interpolate=false, clip=true]{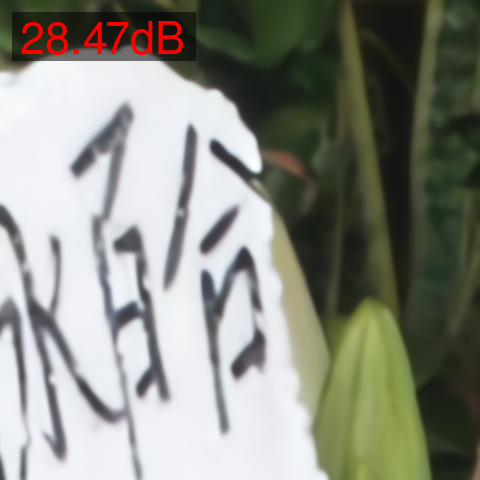}
      \caption{DRN\textsuperscript{*}}
  \end{subfigure}
  \begin{subfigure}[b]{0.16\textwidth}
    \centering
      \includegraphics[width=\textwidth, interpolate=false, clip=true]{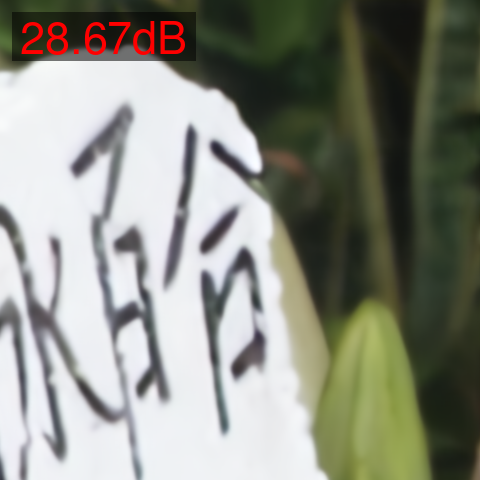}
      \caption{RCAN\textsuperscript{*}}
  \end{subfigure}
  \begin{subfigure}[b]{0.16\textwidth}
    \centering
      \includegraphics[width=\textwidth, interpolate=false, clip=true]{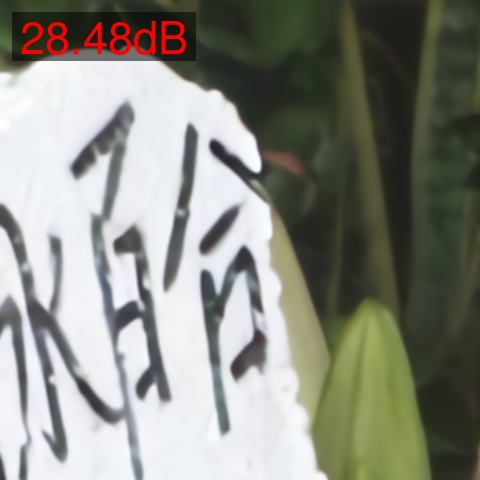}
      \caption{RCAN}
  \end{subfigure}
  \begin{subfigure}[b]{0.16\textwidth}
    \centering
      \includegraphics[width=\textwidth, interpolate=false, clip=true]{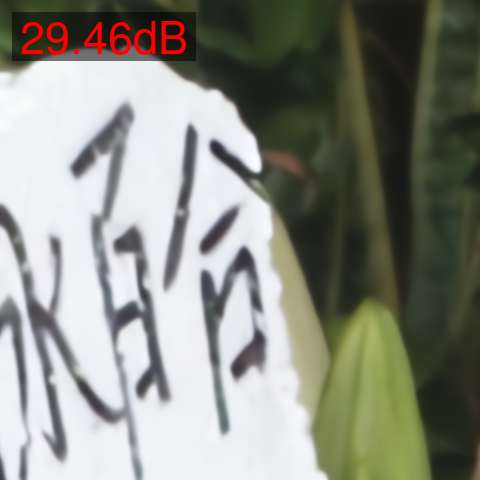}
      \caption{Ensemble}
  \end{subfigure}
  \begin{subfigure}[b]{0.16\textwidth}
    \centering
      \includegraphics[width=\textwidth, interpolate=false, clip=true]{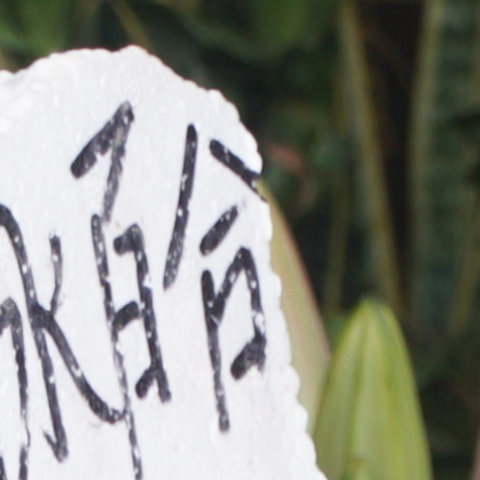}
      \caption{HR}
  \end{subfigure}
\end{center}
    \vspace{-10pt}
    \caption{Visual and quantitative comparison of $\times 4$ SR results.}
    \vspace{-10pt}
\label{fig:x4}
\end{figure}

\subsection{Ablation Study}
\label{sec:as}

Ablation studies were conducted to examine the effectiveness of
the proposed heterogeneous model ensemble. Three models
were trained for all three upscaling factors.
DRN\textsuperscript{*} is the selected DRN model using GP-NAS
where $F=128, D=18, L=3$.  RCAN\textsuperscript{*} is the GP-NAS
selected variation of RCAN~\cite{zhang_eccv_2018} with 128 features,
5 residual groups and 10 residual blocks in each group.
And RCAN uses the original settings in~\cite{zhang_eccv_2018}
with 64 features, 10 residual groups and 20 residual blocks per group.

As shown in Table~\ref{tab:as}, the average performances of three
individual models are very close.  In general, DRN\textsuperscript{*}
is the best with RCAN as the close second.  In comparison, results from
model ensemble is always better than the individual models of the ensemble
no matter it is a 2-model ensemble or 3-model one.
Overall, the best results are always from the 3-model ensemble.
One thing interesting to note is, out of three 2-model ensemble results,
the one combining RCAN and RCAN\textsuperscript{*} is consistently
worse than others.  It could be explained that, for
heterogeneous model ensemble, differences in network architecture
is more beneficial than differences in hyperparameters of
the same network architecture.

Some image examples are shown in Figures~\ref{fig:x2}-\ref{fig:x4},
where PSNR values of SR results are also annotated for quantitative
comparison too.  For small upscaling factor $\times 2$ as in
Fig.\ref{fig:x2}, there is no big difference among SR
results visually.  In the top example, individual SR models
are all able to super-resolve the fine line features which are
blurred in the LR input.  By combining the models together,
the ensemble result has higher PSNR than individual ones.
But the other two examples show that the ensemble PSNR is lower
than the best individual one.  For the middle one, RCAN\textsuperscript{*}
seems to be the outlier comparing with the other two.
Removing outlier before averaging would have increased
the performance of ensemble.  However, the outlier DRN\textsuperscript{*}
in the bottom example is the most accurate and removing it would lead
to worse ensemble result.

Visual difference starts to stand out more in $\times 3$
examples as represented in Fig.\ref{fig:x3}.  For the
first example on the top, all three
individual models are able to super-resolve the small
dot pattern which are blurred in LR input, but each is
slightly different in details.  By combining the models together,
the ensemble result has higher PSNR than individual ones.
The middle example also show ensemble result is higher
than all individual ones and the visual difference
are around the feather-like structures, showing different color
and contrast.  The bottom example shows DRN\textsuperscript{*}
has sharper image than the other two and its PSNR is higher
than the ensemble as a result.

More interesting visual differences are observed in $\times 4$
examples as shown in Fig.\ref{fig:x4}.  For the first two examples,
individual models resolve the blurred details differently,
like the lines near right bottom right corner in the first
and the small text in the second.  While the ensemble
result doesn't increase the image clarity, it has higher
PSNR than all individual ones.  The last one is worth noting
as it demonstrates the difference in the optical transformation
$\mathbb{O}(\cdot)$ for different cameras as explained in
Section~\ref{sec:pf}.  The front object in HR image is clearer
comparing to LR due to higher spatial resolution, but its background
plant is more blurry than the LR counterpart, probably because
the camera used for HR images has a smaller depth-of-field.
The individual SR models handle this quite differently, with
DRN\textsuperscript{*} mimics the depth-of-field of the LR camera
while the other two closer to the HR version. The ensemble method
results in a significant increase in PSNR comparing to individual ones.

\begin{table}[ht!]
\small
\begin{center}
 \begin{tabular}{rcccccc}
 \toprule
     {} & \multicolumn{2}{c}{Upscaling $\times 4$} & \multicolumn{2}{c}{Upscaling $\times 3$} & \multicolumn{2}{c}{Upscaling $\times 2$} \\
     {} & {PSNR} & {SSIM} & {PSNR} & {PSNR} & {PSNR} & {SSIM} \\
 \midrule
 \textbf{Baidu (ours)} & \textbf{31.3960\textsuperscript{1}} & \textbf{0.8751\textsuperscript{1}} & \textbf{30.9496\textsuperscript{1}} & \textbf{0.8762\textsuperscript{1}} & \textbf{33.4460\textsuperscript{1}} & \textbf{0.9270\textsuperscript{1}} \\
 ALONG & 31.2369\textsuperscript{2} & 0.8742\textsuperscript{3} & 30.3745\textsuperscript{5} & 0.8661\textsuperscript{6} & 33.0982\textsuperscript{8} & 0.9238\textsuperscript{6} \\
 CETC-CSKT & 31.1226\textsuperscript{4} & 0.8744\textsuperscript{2} & 30.7651\textsuperscript{2} & 0.8714\textsuperscript{2} & 33.3140\textsuperscript{2} & 0.9245\textsuperscript{2} \\
 SR-IM & 31.1735\textsuperscript{3} & 0.8728\textsuperscript{7} & - & - & - & - \\
 DeepBlueAI & 30.9638\textsuperscript{6} & 0.8737\textsuperscript{4} & 30.3017\textsuperscript{7} & 0.8665\textsuperscript{4} & 33.1771\textsuperscript{7} & 0.9236\textsuperscript{7} \\
 JNSR & 30.9988\textsuperscript{5} & 0.8722\textsuperscript{8} & - & - & - & - \\
 OPPO\_CAMERA & 30.8603\textsuperscript{8} & 0.8736\textsuperscript{5} & 30.5373\textsuperscript{4} & 0.8695\textsuperscript{3} & 33.3091\textsuperscript{3} & 0.9242\textsuperscript{4}  \\
 Kailos & 30.8659\textsuperscript{7} & 0.8734\textsuperscript{6} & 30.1303\textsuperscript{8} & 0.8664\textsuperscript{5} & 32.7084\textsuperscript{12} & 0.9196\textsuperscript{11} \\
 SR\_DL & 30.6045\textsuperscript{9} & 0.8660\textsuperscript{12} & - & - & - & - \\
 Noah\_TerminalVision & 30.5870\textsuperscript{10} & 0.8662\textsuperscript{11} & 30.5641\textsuperscript{3} & 0.8661\textsuperscript{7} & 33.2888\textsuperscript{4} & 0.9228\textsuperscript{8} \\
 Webbzhou & 30.4174\textsuperscript{12} & 0.8673\textsuperscript{10} & - & - & - & - \\ 
 TeamInception & 30.3465\textsuperscript{13} & 0.8681\textsuperscript{9} & - & - & 33.2322\textsuperscript{6} & 0.9240\textsuperscript{5} \\
 lyl & 30.3191\textsuperscript{14} & 0.8655\textsuperscript{13} & 30.3654\textsuperscript{6} & 0.8642\textsuperscript{8} & 32.9368\textsuperscript{10} & 0.9210\textsuperscript{9} \\
 MCML-Yonsei & 30.4201\textsuperscript{11} & 0.8637\textsuperscript{15} & - & - & 32.9032\textsuperscript{11} & 0.9186\textsuperscript{12} \\
 MoonCloud & 30.2827\textsuperscript{15} & 0.8644\textsuperscript{14} & - & - & - & - \\
 qwq & 29.5878\textsuperscript{17} & 0.8547\textsuperscript{16} & 29.2656\textsuperscript{9} & 0.8521\textsuperscript{9} & 31.64\textsuperscript{13} & 0.9126\textsuperscript{13} \\
 SrDance & 29.5952\textsuperscript{16} & 0.8523\textsuperscript{17} & - & - & - & - \\
 MLP\_SR & 28.6185\textsuperscript{18} & 0.8314\textsuperscript{18} & - & - & - & - \\
 RRDN\_IITKGP & 27.9708\textsuperscript{19} & 0.8085\textsuperscript{20} & - & - & 29.8506\textsuperscript{14} & 0.8453\textsuperscript{14} \\
 congxiaofeng & 26.3915\textsuperscript{20} & 0.8258\textsuperscript{19} & - & - & - & - \\
 AiAiR & - & - & 18.1903\textsuperscript{10} & 0.8245\textsuperscript{10} & 33.2633\textsuperscript{5} & 0.9243\textsuperscript{3} \\
 GDUT-SL & - & - & - & - & 32.9725\textsuperscript{9} & 0.9204\textsuperscript{10} \\
 \bottomrule
 \end{tabular}
\end{center}
     \caption{Quantitative results of all three tracks for the AIM 2020 Real Image Super-Resolution Challenge. The superscript number indicates ranking of each metric.}
     \vspace{-10pt}
 \label{tab:aimsr}
\end{table}

\subsection{AIM 2020 Challenge Results}
\label{sec:aimc}

To generate the full-size SR images for the AIM 2020 challenge, all three
models from the ablation study were used for all three upscaling tracks.
For $\times 4$ track, a double regression model~\cite{guo_cvpr_2020} was also trained
to include in the model ensemble.  Each full-size LR test images were
cropped to $120\times 120$ patches and self-ensemble ($\times 8$) was applied.
The cropping window was slided at 60-pixel spacing and the overlapping
patches were average using weights correlated with the distance between
the patch center and each pixel.  With ensemble applied to all three levels,
the generated full-size images were submitted to the challenge and won
the first place at all three tracks.  As shown in Table~\ref{tab:aimsr},
both our PSNR and SSIM values lead the second place with a comfortable margin.

\begin{figure}[ht!]
\captionsetup[subfigure]{labelformat=empty}
\begin{center}
  \begin{subfigure}[b]{0.48\textwidth}
    \centering
      \includegraphics[width=\textwidth, interpolate=true]{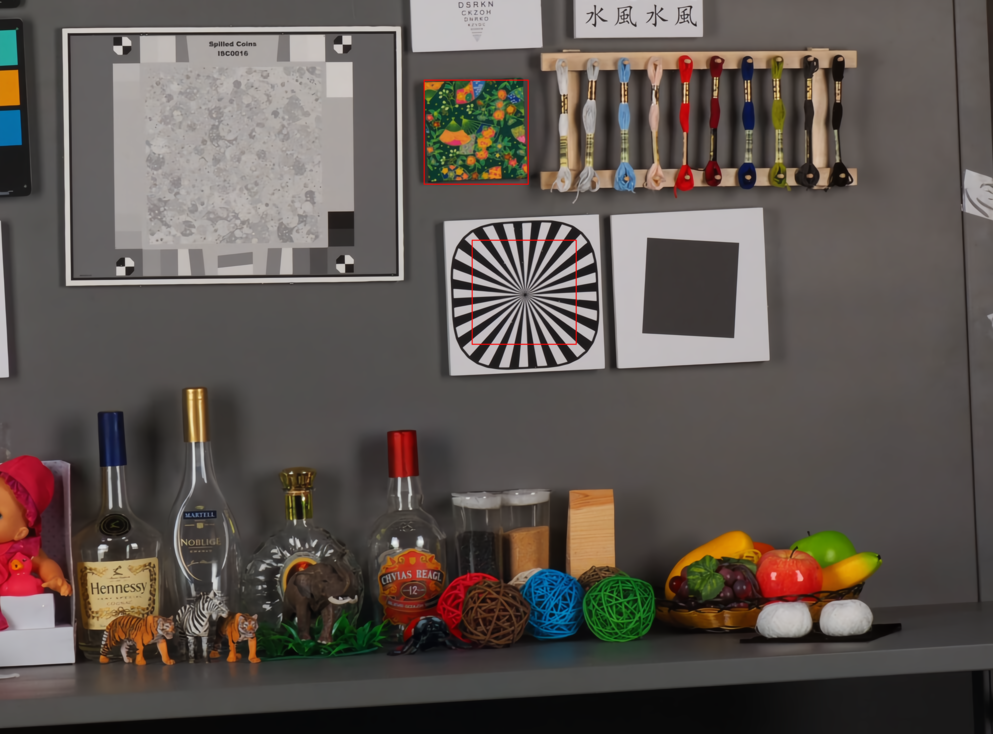}
  \end{subfigure}
  \begin{subfigure}[b]{0.48\textwidth}
    \centering
      \includegraphics[width=\textwidth, interpolate=true]{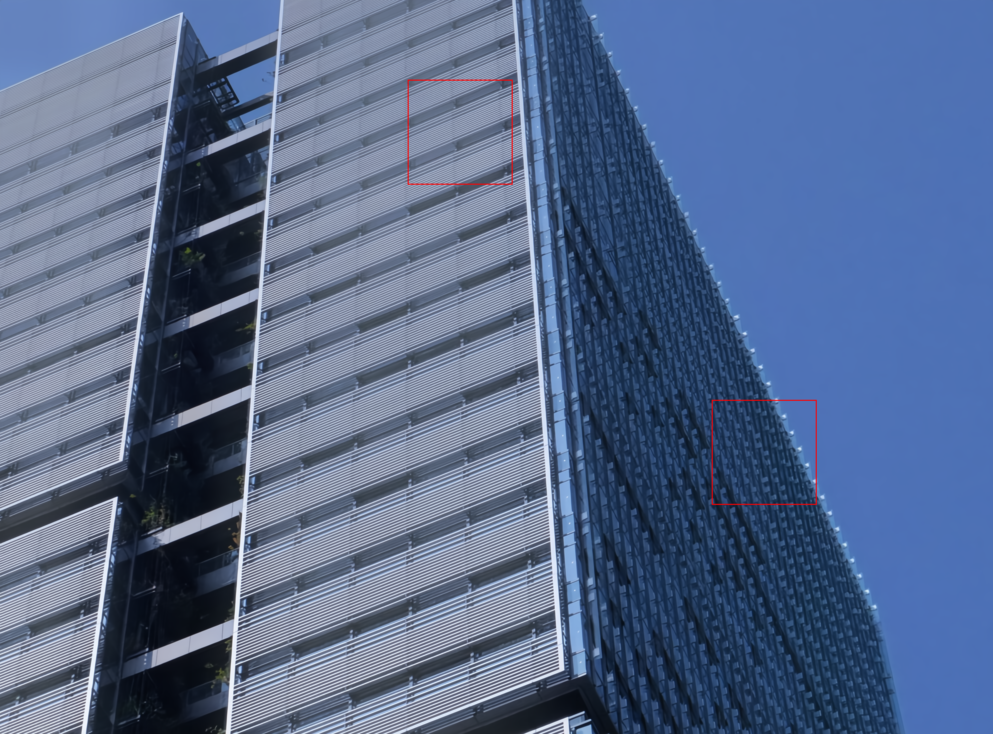}
  \end{subfigure}
  \begin{subfigure}[b]{0.24\textwidth}
    \centering
      \vspace{-0.25em}
      \includegraphics[width=\textwidth, interpolate=false]{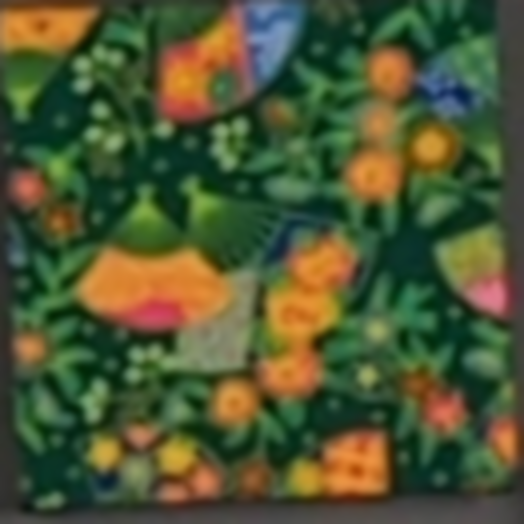}
  \end{subfigure} \hspace*{-0.5em}
  \begin{subfigure}[b]{0.24\textwidth}
    \centering
      \vspace{-0.25em}
      \includegraphics[width=\textwidth, interpolate=false]{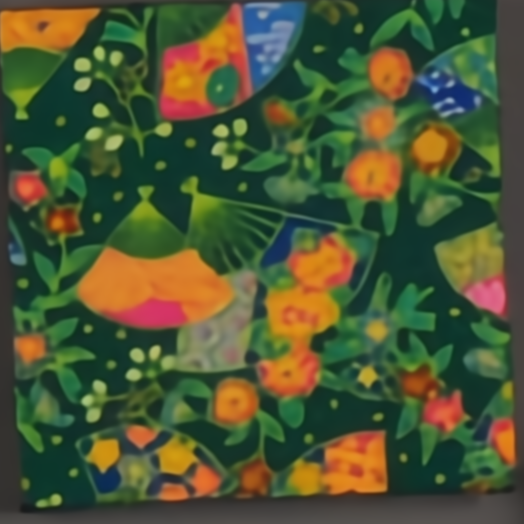}
  \end{subfigure}
  \begin{subfigure}[b]{0.24\textwidth}
    \centering
      \vspace{-0.25em}
      \includegraphics[width=\textwidth, interpolate=false]{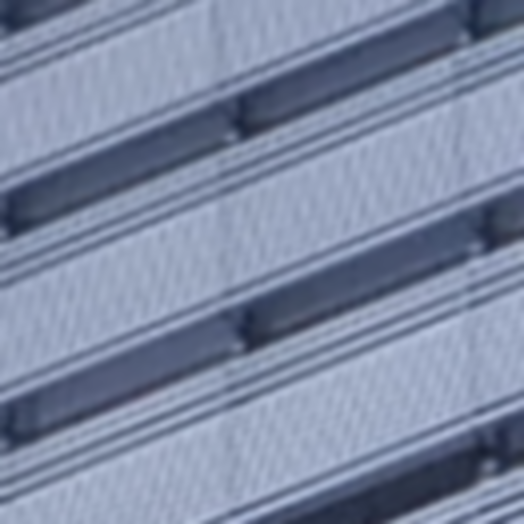}
  \end{subfigure} \hspace*{-0.5em}
  \begin{subfigure}[b]{0.24\textwidth}
    \centering
      \vspace{-0.25em}
      \includegraphics[width=\textwidth, interpolate=false]{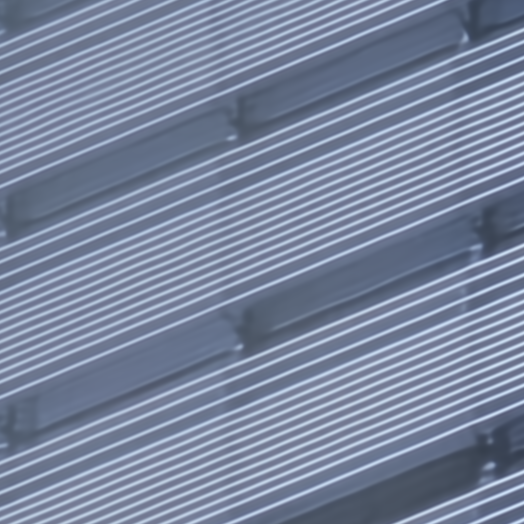}
  \end{subfigure}
  \begin{subfigure}[b]{0.24\textwidth}
    \centering
      \vspace{-0.25em}
      \includegraphics[width=\textwidth, interpolate=false]{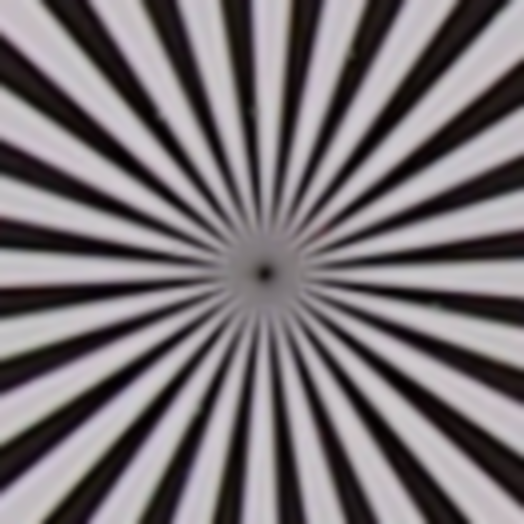}
      \caption{Bicubic}
  \end{subfigure} \hspace*{-0.5em}
  \begin{subfigure}[b]{0.24\textwidth}
    \centering
      \vspace{-0.25em}
      \includegraphics[width=\textwidth, interpolate=false]{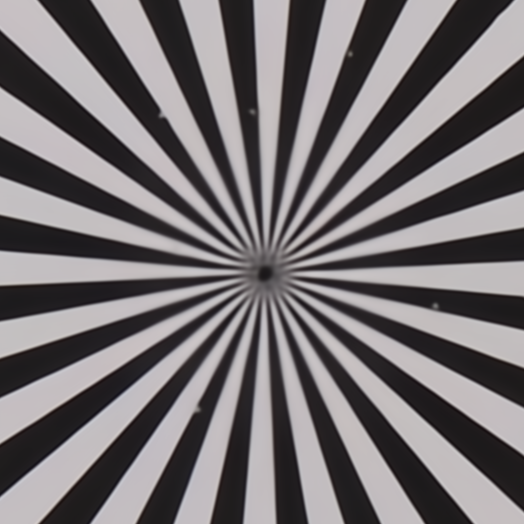}
      \caption{SR\_Ensaemble}
  \end{subfigure}
  \begin{subfigure}[b]{0.24\textwidth}
    \centering
      \vspace{-0.25em}
      \includegraphics[width=\textwidth, interpolate=false]{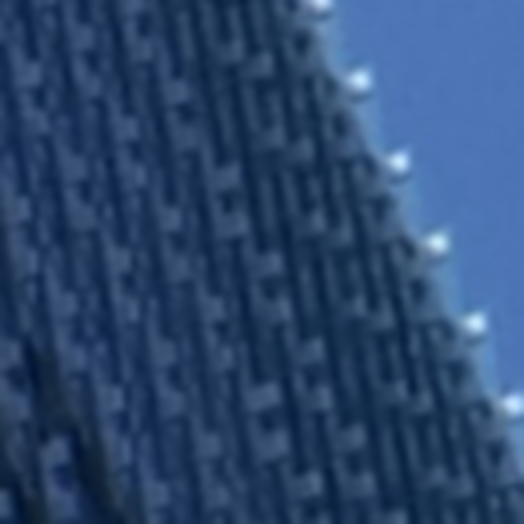}
      \caption{Bicubic}
  \end{subfigure} \hspace*{-0.5em}
  \begin{subfigure}[b]{0.24\textwidth}
    \centering
      \vspace{-0.25em}
      \includegraphics[width=\textwidth, interpolate=false]{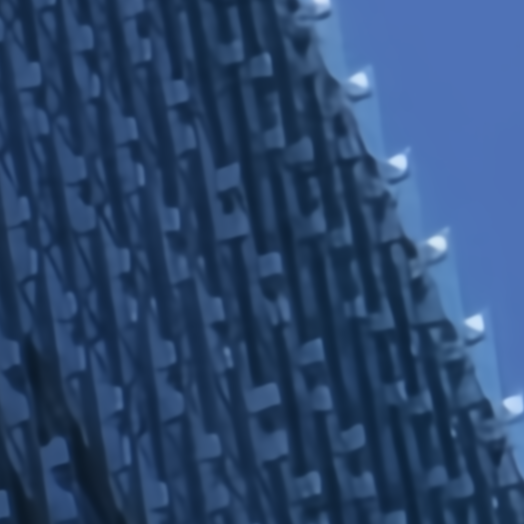}
      \caption{SR\_Ensaemble}
  \end{subfigure}
 \end{center}
    \vspace{-10pt}
    \caption{Examples of full-size test images from the AIM 2020 challenge ($\times 4$).}
    \vspace{-10pt}
\label{fig:fs_x4}
\end{figure}

One set of representative $\times 4$ images are shown in the
Fig.~\ref{fig:fs_x4}.  The full-size SR output are located at
the top, with selected areas zoomed in to compare with
bicubic interpolation results.

\section{Conclusions}
In this paper, based on the models searched via GP-NAS, we have introduced a new heterogeneous model
ensemble method for real image super resolution. Since network architecture greatly affects the results
of SR, we first apply GP-NAS approach to search the key factors such as the number of residual network 
block, block size and the number of features in our network structure.  Then, different models selected 
using GP-NAS are fused together to boost the performance of SR.
Combined with patch-ensemble and self-ensemble, the proposed new scheme is validated to be
highly effective, generating impressive testing results on all three 
tracks ($\times 2$, $\times 3$ and $\times 4$) of the AIM 2020 challenge in terms of both PSNR
and SSIM.

\bibliographystyle{unsrt}  


\end{document}